\documentclass[aps,pra,twocolumn,showpacs,superscriptaddress,floatfix,nofootinbib]{revtex4}

\usepackage{graphicx}
\usepackage{amsmath}
\usepackage{epsfig}
\usepackage{helvet}
\usepackage{amssymb}

\newcommand{\be}{\begin{equation}}
\newcommand{\ee}{\end{equation}}
\newcommand{\bea}{\begin{eqnarray}}
\newcommand{\eea}{\end{eqnarray}}

\newcommand{\tr}{\mbox{tr}}
\newcommand{\bra}[1]{\mbox{$\langle #1 |$}}
\newcommand{\ket}[1]{\mbox{$| #1 \rangle$}}

\def\tr{ \mbox{tr}}

\begin{document}

\title{Entanglement renormalization, scale invariance, and quantum criticality}
\author{Robert N. C. Pfeifer}
\affiliation{School of Physical Sciences, the University of
Queensland, QLD 4072, Australia}
\author{Glen Evenbly}
\affiliation{School of Physical Sciences, the University of
Queensland, QLD 4072, Australia}
\author{Guifr\'e Vidal}
\affiliation{School of Physical Sciences, the University of
Queensland, QLD 4072, Australia}
\date{April 10, 2009}

\begin{abstract}
The use of entanglement renormalization in the presence of scale invariance is investigated. We explain how to compute an accurate approximation of the critical ground state of a lattice model, and how to evaluate local observables, correlators and critical exponents. Our results unveil a precise connection between the multi-scale entanglement renormalization ansatz (MERA) and conformal field theory (CFT). Given a critical Hamiltonian on the lattice, this connection can be exploited to extract most of the conformal data of the CFT that describes the model in the continuum limit.
\end{abstract}

\pacs{03.67.--a,~05.50.+q,~11.25.Hf~~~~~~~~~~~~~~~~~~~~~~\textbf{Published~as}:~Phys.~Rev.~A~79(4),~040301(R)~(2009)\\(C) American Physical Society (2009)}

\maketitle

The study of quantum critical phenomena through real-space renormalization group (RG) techniques \cite{Wilson,DMRG} has traditionally been obstructed by the accumulation, over successive RG transformations, of short-range entanglement across block boundaries. Entanglement renormalization \cite{ER} was recently proposed as a technique to address this problem. By removing short-range entanglement at each iteration of the RG transformation, not only can arbitrarily large lattice systems be considered, but the scale invariance characteristic of critical phenomena is also seen to be restored \cite{ER,Free}. 

In this paper we explain how to use the \emph{multi-scale entanglement renormalization ansatz} (MERA) \cite{MERA} to investigate scale invariant systems \cite{ER,MERA,Free,Topological,Transfer}. It has been showed that the \emph{scale invariant} MERA can represent the infra-red limit of topologically ordered phases \cite{Topological}. Here we focus instead on its use at quantum criticality. We present the following results: (i) given a critical Hamiltonian, an adaptation of the algorithm of Ref. \cite{MERAalgorithmNEW} to compute a scale invariant MERA for its ground state; then, starting from a scale invariant MERA, (ii) a procedure to identify the scaling operators/dimensions of the theory and (iii) a closed expression for two-point and three-point correlators; (iv) a connection between the MERA and conformal field theory, which can be used to readily identify the continuum limit of a critical lattice model; finally (v) benchmark calculations for the Ising and Potts models. 

\begin{figure}[!tb]
\begin{center}
\includegraphics[width=8cm]{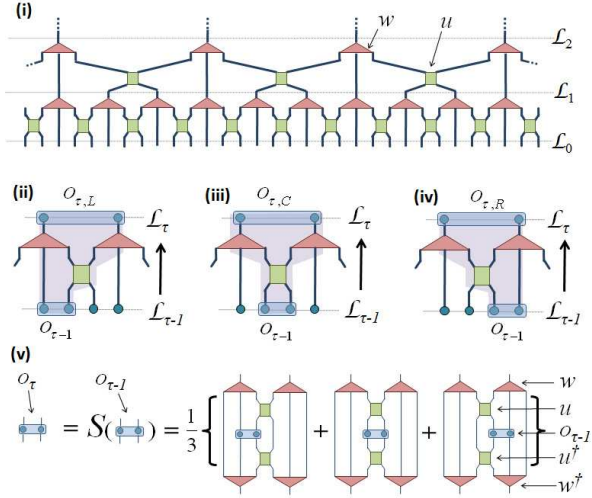}
\caption{(Color online) 
($i$) Two lowest rows of disentanglers $u$ and isometries $w$ of the ternary MERA. They map the original infinite lattice $\mathcal{L}_0\equiv \mathcal{L}$ into increasingly coarse-grained lattices $\mathcal{L}_{1}$ and $\mathcal{L}_2$. Notice that three sites of $\mathcal{L}_{\tau-1}$ become one site of $\mathcal{L}_{\tau}$, hence the use of $\log_3$ throughout the paper. ($ii$)-($iv$) Under the coarse-graining transformation defined by the MERA, two-site operators supported on three different pairs of sites of $\mathcal{L}_{\tau-1}$ become supported on the same pair of sites of $\mathcal{L}_{\tau}$. ($v$) Accordingly, the scaling superoperator $\mathcal{S}$ is the average of three contributions, each of which (and thus also their average) is unital and contractive thanks to the isometric character of $u$ and $w$ \cite{MERA}.} 
\label{fig:TwoSiteCFT}
\end{center}
\end{figure}

We note that result (ii) was already discussed by Giovannetti, Montangero and Fazio in Ref. \cite{Transfer} using the \emph{binary} MERA of Ref. \cite{MERA}. Our derivations are conducted instead with the \emph{ternary} MERA of Ref \cite{MERAalgorithmNEW} (see Fig. \ref{fig:TwoSiteCFT}), in terms of which results (iii)-(iv) acquire a simple form.

We start by considering a finite 1D lattice $\mathcal{L}$ made of $N$ sites, each one described by a vector space $\mathbb{V}$ of dimension $\chi$. The (ternary) MERA is a tensor network that serves as an ansatz for pure states $\ket{\Psi}\in \mathbb{V}^{\otimes N}$ of the lattice, see Fig. \ref{fig:TwoSiteCFT}. Its tensors, known as \emph{disentanglers} and \emph{isometries}, are organized in $T\approx \log_3 N$ layers, each one implementing a RG transformation. Such transformations produce a sequence of lattices,
\begin{equation}
\mathcal{L}_0  ~\rightarrow  ~\mathcal{L}_1 ~\rightarrow ~\cdots ~\rightarrow ~\mathcal{L}_T, ~~~~~~~~~\mathcal{L}_{0} \equiv \mathcal{L},
\end{equation}
where lattice $\mathcal{L}_{\tau+1}$ is a coarse-graining of lattice $\mathcal{L}_{\tau}$, and the top lattice $\mathcal{L}_T$ is sufficiently small to allow exact numerical computations. Let $o$ denote a local observable supported on two contiguous sites of $\mathcal{L}$, and let $\rho_{T}$ be the density matrix that describes the state of the system on two contiguous sites of $\mathcal{L}_{T}$. Then the \emph{ascending} and \emph{descending} superoperators $\mathcal{A}_{\tau}$ and $\mathcal{D}_{\tau}$ \cite{MERAalgorithmNEW}, 
\begin{equation}
	o_{\tau} = \mathcal{A}_{\tau}(o_{\tau-1}),~~~~~~~~~~ \rho_{\tau-1} = \mathcal{D}_{\tau}(\rho_{\tau}),
\end{equation}
generate a sequence of operators and density matrices 
\begin{eqnarray}
	&&o_0  \stackrel{\mathcal{A}_1}{\rightarrow}  o_1 \stackrel{\mathcal{A}_2}{\rightarrow} ~\cdots~ \stackrel{\mathcal{A}_T}{\rightarrow} o_T, ~~~~~~~~~ o_0 \equiv o, \label{eq:A}\\
	&&\rho_0  \stackrel{\mathcal{D}_1}{\leftarrow}  \rho_1 \stackrel{\mathcal{D}_2}{\leftarrow} ~\cdots~ \stackrel{\mathcal{D}_T}{\leftarrow} \rho_T, ~~~~~~~~~ \rho_0 \equiv \rho, \label{eq:D}
\end{eqnarray}
where $o_{\tau}$ and $\rho_{\tau}$ are supported on two contiguous sites of the lattice $\mathcal{L}_{\tau}$. Eq. (\ref{eq:A}) allows us to monitor how the local operator $o$ transforms under successive RG transformations, whereas its expected value $\left\langle o \right\rangle = \tr(\rho o)$ can be evaluated by computing $\rho$ in Eq. (\ref{eq:D}).

%%%%%%%%%%%%%%%%%%%%%%%%%%%%%%%%%%%%%%%%%%%%%%%%%%%%%%%%%%%%%%%%%%%%%%%%%%%%%%%%%%%%%%%%%%%%%%%%%

\textbf{RG fixed point.---} The scale invariant MERA corresponds to the limit of infinitely many layers, $T\rightarrow \infty$, and to choosing the disentanglers and isometries in all layers to be copies of a unique pair $u$ and $w$ \cite{ER,MERA}. In this case we refer to the ascending superoperator $\mathcal{A}_{\tau}$, which no longer depends on $\tau$, as the \emph{scaling superoperator} $\mathcal{S}$ (see Fig. \ref{fig:TwoSiteCFT}), and to its dual $\mathcal{D}_{\tau}$ as $\mathcal{S}^{*}$. Notice that $\mathcal{S}$ is a fixed-point RG map. Then, as customary in RG analysis \cite{Cardy,Francesco}, the scaling operators $\phi_{\alpha}$ and scaling dimensions $\Delta_{\alpha}$ of the theory,
\begin{equation}
	\mathcal{S}(\phi_{\alpha}) = \lambda_{\alpha} \phi_{\alpha},~~~~~~~\Delta_{\alpha} \equiv -\log_3 \lambda_{\alpha},
	\label{eq:scaling}
\end{equation}
are obtained by diagonalizing this map,
\begin{equation}
	\mathcal{S}(\bullet) = \sum_{\alpha} \lambda_{\alpha} \phi_{\alpha} \tr(\hat{\phi}_{\alpha} \bullet),~~~~~~\tr(\hat{\phi}_{\alpha} \phi_{\beta}) = \delta_{\alpha\beta},
	\label{eq:spectral}
\end{equation}
where $\hat{\phi}_{\alpha}$ are the eigenvectors of the dual $S^{*}$, $S^{*}(\hat{\phi}_{\alpha}) = \lambda_{\alpha} \hat{\phi}_{\alpha}$. Eq. \ref{eq:spectral} was first discussed in Ref. \cite{Transfer} by Giovannetti, Montangero and Fazio \cite{clarify}. It formalizes a previous observation (see Eq. 5 of Ref. \cite{MERA}) that the scale invariant MERA displays polynomial correlations. By construction, $\mathcal{S}$ is \emph{unital}, $\mathcal{S}(\mathbb{I}) = \mathbb{I}$, so that the identity operator $\mathbb{I}$ in $\mathbb{V}^{\otimes 2}$ is a scaling operator with eigenvalue $\lambda_{\mathbb{I}}=1$; and \emph{contractive}, meaning $|\lambda_{\alpha}|\leq 1$ \cite{Super}. Here we will assume, as it is the case in the examples below, that only the identity operator $\mathbb{I}$ has eigenvalue $\lambda = 1$. Then the operator $\hat{\rho} \equiv \hat{\mathbb{I}}$ is a density matrix that corresponds to the \emph{unique} fixed point of $\mathcal{S}^{*}$, $\mathcal{S}^{*}(\hat{\rho})=\hat{\rho}$, and since
\begin{equation}
 \lim_{T\rightarrow \infty} \big(\underbrace{\mathcal{S}^{*}\circ \cdots \circ \mathcal{S}^{*}}_{T \mbox{ \scriptsize{times}}}\big) (\rho_T) = \hat{\rho}
	\label{eq:rho}
\end{equation}
for any starting $\rho_T$, it follows that $\hat{\rho}$ is the state of any pair of contiguous sites of $\mathcal{L}$. [Consistent with scale invariance, $\hat{\rho}$ is also the state of any pair of contiguous sites of $\mathcal{L}_{\tau}$ for any finite $\tau$]. The computation of the expected value of the local observable $o$ is then straightforward,
\begin{equation}
\left\langle o \right\rangle = \tr(\hat{\rho} o),
\end{equation}
which for the scaling operators reduces to	$\left\langle \phi_{\alpha} \right\rangle = \delta_{\alpha \mathbb{I}}$.

%%%%%%%%%%%%%%%%%%%%%%%%%%%%%%%%%%%%%%
%%%%%%%%%%%%%%%%%%%%%%%%%%%%%%%%%%%%%%
%%%%%%%%%%%%%%%%%%%%%%%%%%%%%%%%%%%%%
%%%%%%%%%%%%%%%%%%%%%%%%%%%%%%%%%%%%%

%%%%%%%%%%%%%%%%%%%%%%%%%%%%%%%%%%%%%%%%%%%%%%%%%%%%%%%%%%%%%%%%%%%%%%%%%%%%%%%%%%%%%%%%%%%%%%%%%%

\begin{figure}[!tb]
\begin{center}
\includegraphics[width=9cm]{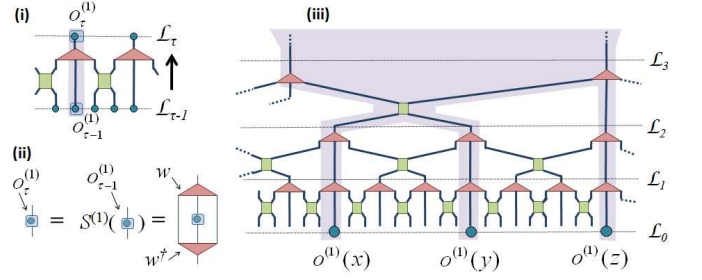}
\caption{(Color online) 
($i$) One-site operators on special sites are coarse-grained into one-site operators. ($ii$) Scaling superoperator for one-site operators. ($iii$) In computing correlators on specific sites $x$ and $y$ (or $x$, $y$ and $z$), one-site operators are coarse-grained individually according to $\mathcal{S}^{(1)}$ until they become nearest neighbors (which in this case occurs at lattice $\mathcal{L}_{2}$, $q=2$).} 
\label{fig:OneSiteCFT}
\end{center}
\end{figure}

\textbf{Correlators.---} Let us now diagonalize the one-site scaling superoperator $\mathcal{S}^{(1)}$ of Fig. \ref{fig:OneSiteCFT}, 
\begin{equation}
	\mathcal{S}^{(1)}(\bullet) = \sum_{\alpha} \lambda^{(1)}_{\alpha} \phi^{(1)}_{\alpha} \tr(\hat{\phi}^{(1)}_{\alpha}\bullet), \label{eq:spectral1}
\end{equation}
where the scaling dimensions $\Delta^{(1)}_{\alpha}\equiv -\log_3 \lambda^{(1)}_{\alpha}$ coincide with $\Delta_{\alpha}$ \cite{ScDim}. The correlator for two scaling operators $\phi_{\alpha}^{(1)}$ and $\phi_{\beta}^{(1)}$ placed on contiguous sites reads
\begin{equation}
	C_{\alpha\beta} \equiv \left\langle \phi_{\alpha}^{(1)}(1) \phi_{\beta}^{(1)}(0)\right\rangle 
	= \tr \big( (\phi^{(1)}_{\alpha}\otimes \phi^{(1)}_{\beta}) \hat{\rho} \big).
	\label{eq:C2}
\end{equation}
Suppose now that $\phi_{\alpha}^{(1)}$ and $\phi_{\beta}^{(1)}$ are placed in two special sites $x,y$ as in Fig. \ref{fig:OneSiteCFT}, where $r_{xy}\equiv x-y$ is such that $|r_{xy}| = 3^q$ for $q=1,2,\cdots$. Then after $q = \log_3 |r_{xy}|$ iterations of the RG transformation, $\phi_{\alpha}^{(1)}$ and $\phi_{\beta}^{(1)}$ become first neighbors again. Notice that each iteration contributes a factor $\lambda^{(1)}_{\alpha}\lambda^{(1)}_{\beta}$. Using the identity $a^{\log b}=b^{\log a}$ we find 
\begin{equation}
(\lambda^{(1)}_{\alpha}\lambda^{(1)}_{\beta})^{\log_3 |r_{xy}|} = |r_{xy}|^{\log_3 (\lambda^{(1)}_{\alpha}\lambda^{(1)}_{\beta})} = |r_{xy}|^{-\Delta^{(1)}_{\alpha}-\Delta^{(1)}_{\beta}}
\nonumber
\end{equation}
and obtain a closed expression for two-point correlators,
\begin{equation}
	\left\langle \phi_{\alpha}^{(1)}(x) \phi_{\beta}^{(1)}(y)\right\rangle 
	= \frac{C_{\alpha\beta}}{|r_{xy}|^{\Delta^{(1)}_{\alpha}+\Delta^{(1)}_{\beta}}}.
	\label{eq:two-point}
\end{equation}
For three-point correlators we define the constants
\begin{eqnarray}
	\Omega_{\alpha\beta}^{~\gamma} &\equiv& \Delta^{(1)}_{\alpha}+\Delta^{(1)}_{\beta}-\Delta^{(1)}_{\gamma}\\
	C_{\alpha\beta\gamma} &\equiv& 2^{\Omega_{\gamma\alpha}^{~\beta}}\tr \big( (\phi^{(1)}_{\alpha}\otimes \phi^{(1)}_{\beta}\otimes \phi^{(1)}_{\gamma}) \hat{\rho}^{(3)} \big) 
\label{eq:C3}
\end{eqnarray}
where the trace corresponds to the correlator on three consecutive sites and $\hat{\rho}^{(3)}$ is obtained from $\hat{\rho}$. For $|r_{xy}| = |r_{yz}| = |r_{xz}|/2 = 3^q$, analogous manipulations lead to
\begin{eqnarray}
	\left\langle \phi_{\alpha}^{(1)}(x) \phi_{\beta}^{(1)}(y) \phi_{\beta}^{(1)}(z)\right\rangle 
	= \frac{C_{\alpha\beta\gamma}}{
	|r_{xy}|{}^{\Omega_{\alpha\beta}^{~\gamma}}
	|r_{yz}|{}^{\Omega_{\beta\gamma}^{~\alpha}}
	|r_{zx}|{}^{\Omega_{\gamma\alpha}^{~\beta}}
	} \label{eq:three-point}
\end{eqnarray}

%%%%%%%%%%%%%%%%%%%%%%%%%%%%%%%%%%%%%%
%%%%%%%%%%%%%%%%%%%%%%%%%%%%%%%%%%%%%%
%%%%%%%%%%%%%%%%%%%%%%%%%%%%%%%%%%%%%
%%%%%%%%%%%%%%%%%%%%%%%%%%%%%%%%%%%%%

%%%%%%%%%%%%%%%%%%%%%%%%%%%%%%%%%%%%%%%%%%%%%%%%%%%%%%%%%%%%%%%%%%%%%%%%%%%%%%%%%%%%%%%%%%%%%%%%%%%

\textbf{CFT.---} The continuous limit of a quantum criticial lattice system (scale invariant case) corresponds to a conformal field theory (CFT) \cite{Cardy,Francesco}. A CFT contains an infinite set of quasi-primary fields $\phi^{\mbox{\tiny CFT}}_{\alpha}$, with scaling dimensions $\Delta_{\alpha}^{\mbox{\tiny CFT}}$. The correlators involving two or three quasi-primary fields have expressions analogous to Eqs. \ref{eq:two-point} and \ref{eq:three-point}, and the (symmetric) coefficients $C_{\alpha\beta\gamma}^{\mbox{\tiny CFT}}$ for three-point correlators coincide with those in the so-called \emph{operator product expansion} (OPE). Moreover, quasi-primary fields are organized in \emph{conformal towers} corresponding to irreducible representations of the Virasoro algebra. Each tower contains one \emph{primary field} $\phi^{p}$ at the top, with conformal dimensions $(t,\bar{t})$ [such that its scaling dimension is $\Delta^{p} \equiv t + \bar{t}\,$], and its infinitely many descendants, which are quasi-primary fields with scaling dimension $\Delta = \Delta^{p} + n$ for some integer $n\geq 1$. 

A CFT is completely specified by its symmetries once the following conformal data has been provided: (i) the \emph{central charge} $c$, (ii) a complete list of \emph{primary fields} with their \emph{conformal dimensions} and (iii) the OPE for these primary fields. For instance, the Ising CFT in 1+1 dimensions has central charge $c=1/2$, three primary fields \emph{identity} $\mathbb{I}$, \emph{spin} $\sigma$ and \emph{energy} $\epsilon$ with conformal dimensions $(0,0)$, $(\frac{1}{16},\frac{1}{16})$ and $(\frac{1}{2},\frac{1}{2})$, and OPE coefficients
\begin{eqnarray}
	C^{\mbox{\tiny CFT}}_{\alpha\beta \mathbb{I}}\!= \!\delta_{\alpha\beta}, ~
  C^{\mbox{\tiny CFT}}_{\sigma\sigma\epsilon} \! = \frac{1}{2}, ~~
	C^{\mbox{\tiny CFT}}_{\sigma\sigma\sigma}\! = \!C^{\mbox{\tiny CFT}}_{\epsilon\epsilon\epsilon}\! =\! C^{\mbox{\tiny CFT}}_{\epsilon\epsilon\sigma}\! = 0.
	\label{eq:OPE_Ising}
\end{eqnarray}

The present analysis readily suggests a correspondence between the scaling operators $\phi_{\alpha}$ of the scale invariant MERA, defined on a lattice, and the quasi-primary fields $\phi_{\alpha}^{\mbox{\tiny CFT}}$ of a CFT, defined in the continuum. Together with the algorithm described below, this correspondence grants us numerical access, given a critical Hamiltonian $H$ on the lattice, to most of the conformal data of the underlying CFT, namely to scaling dimensions and OPE coefficients. The central charge $c$ can also be obtained e.g. \cite{Latorre} from the von Neumann entropy $S(\rho)\equiv-\tr(\rho\log_2\rho)$, which for a block of $L$ sites scales, up to some additive constant, as $S = \frac{c}{3} \log_2 L$ \cite{entropy}. We then have $S(\hat{\rho}) - S(\hat{\rho}^{(1)}) =  \frac{c}{3} (\log_2 2-\log_2 1) = \frac{c}{3}$, or simply
\begin{equation}
	c = 3 \left(S(\hat{\rho}) - S(\hat{\rho}^{(1)})\right).
	\label{eq:central}
\end{equation}

%%%%%%%%%%%%%%%%%%%%%%%%%%%%%%%%%%%%%%%%%%%%%%%%%%%%%%%%%%%%%%%%%%%%%%%%%%%%%%%%%%%%%%%%%%%%%%%%%

\textbf{Algorithm.---} Given a critical Hamiltonian $H$ for an infinite lattice, we obtain a scale invariant MERA for its ground state $\ket{\Psi}$ by adapting the general strategy discussed in Ref. \cite{MERAalgorithmNEW}. Recall that tensors (disentanglers $u$ and isometries $w$) are optimized so as to minimize the energy $E \equiv \bra{\Psi} H \ket{\Psi}$. After linearization this reads 
\begin{equation}
	E = \tr(u \Upsilon_u) + k_1 = \tr(w \Upsilon_w) + k_2,
\end{equation}
where $\Upsilon_u$ and $\Upsilon_{w}$ are known as \emph{environments} and $k_1,k_2$ are two irrelevant constants. In the translation invariant case \cite{MERAalgorithmNEW} the environment for, say, an isometry $w$ at layer $\tau$ of the MERA, $\Upsilon_w = f(u_{\tau},w_{\tau},\rho_{\tau}, h_{\tau-1})$, is a function of the disentangler $u_{\tau}$ and isometry $w_{\tau}$ of that layer, a two-site density matrix $\rho_{\tau}$ and a two-site Hamiltonian term $h_{\tau-1}$. In the present case, we replace the above with the unique pair $(u,w)$, the fixed-point density matrix $\hat{\rho}$, and an average Hamiltonian $\bar{h} \equiv \sum_{\tau} h_{\tau}/3^{\tau}$, where the weights $1/3^{\tau}$ account for the relative number of tensors in different layers of the MERA. Then, starting from some initial pair $(u,w)$ and the critical Hamiltonian $H$ made of two-body terms $h$, the following steps are repeated until convergence:
\newcounter{Lcount} 
\begin{list}{A\arabic{Lcount}.}  {\usecounter{Lcount}}\setcounter{Lcount}{0} 
	\item Given the latest $(u,w)$, compute $(\hat{\rho},\bar{h})$.
	\item Given $(u,w,\hat{\rho},\bar{h})$, update the pair $(u,w)$.
\end{list}
In step A1, the scaling superoperator $\mathcal{S}$ is built as indicated in Fig. \ref{fig:TwoSiteCFT}. We compute the fixed-point density matrix $\hat{\rho}$ by sparse diagonalization of $\mathcal{S}$, and the average Hamiltonian $\bar{h}$ by using $h_{\tau} = \mathcal{S}(h_{\tau-1})$, $h_0\equiv h$ \cite{truncation}. Step A2 is decomposed into a sequence of alternating optimizations for $u$ and $w$ as in the generic algorithm of Ref. \cite{MERAalgorithmNEW}, where each tensor is updated by computing a singular value decomposition of its environment.

%%%%%%%%%%%%%%%%%%%%%%%%%%%%%%%%%%%%%%
%%%%%%%%%%%%%%%%%%%%%%%%%%%%%%%%%%%%%%
\begin{figure}[!tb]
\begin{center}
\includegraphics[width=8cm]{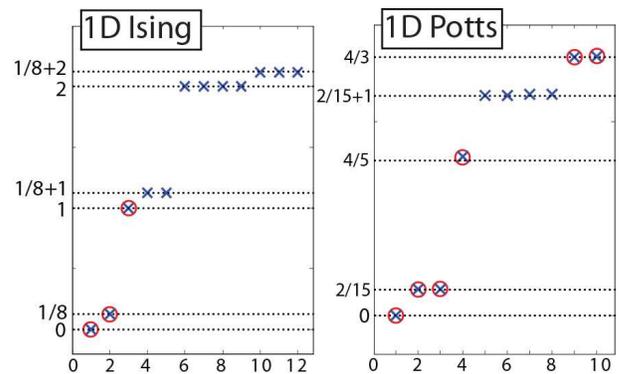}
\caption{(Color online) 
Scaling dimensions $\Delta_{\alpha}$ obtained from the spectrum of the scaling superoperator $\mathcal{S}$. Circles indicate primary fields. \emph{Left:} For the Ising model we can identify the scaling dimensions of the three primary fields, the so-called identity $\mathbb{I}$, spin $\sigma$ and energy $\epsilon$, together with several of their descendants. \emph{Right:} The spectrum of $\mathcal{S}$ for the 3-level Potts model shows some of its primary fields, including its primary fields with multiplicity two, namely the spins $\sigma_1$ and $\sigma_2$ and the pair $Z_1$ and $Z_2$ \cite{Francesco}. 
} 
\label{fig:CFTscaling}
\end{center}
\end{figure}
%%%%%%%%%%%%%%%%%%%%%%%%%%%%%%%%%%%%%
%%%%%%%%%%%%%%%%%%%%%%%%%%%%%%%%%%%%%

\textbf{Examples.---} We illustrate the above ideas and the performance of the algorithm by considering the Ising and 3-level Potts quantum critical models in 1D,
\begin{eqnarray}
H_{{\rm{Ising}}}  &=& \sum_r \left(\lambda \sigma^{[r]}_z  + \sigma^{[r]}_x \sigma^{[r+1]}_x \right) \\
%H_{{\rm{Potts}}}  &=& \sum_r \left(\lambda M^{[r]}_{z}  + \sum_{a=1,2} M^{[r]}_{x,a} M^{[r+1]}_{x,3-a} \right)
H_{{\rm{Potts}}}  &=& \sum_r \left(\lambda M^{[r]}_{z}  +  M^{[r]}_{x,1} M^{[r+1]}_{x,2} + M^{[r]}_{x,2} M^{[r+1]}_{x,1}\right) \nonumber
 \label{hams}
\end{eqnarray}
where $\sigma_z$ and $\sigma_x$ are Pauli matrices, and
%whereas $M_z$, $M_{x,1}$ and $M_{x,2}$ are given by
\begin{eqnarray}
	M_{z} = \left( \begin{array}{ccc} 
		2 & 0 & 0\\
		0 & -1 & 0\\
		0 & 0 & -1
	\end{array} \right), M_{x,1}	= \left( \begin{array}{ccc} 
		0 & 1 & 0\\
		0 & 0 & 1\\
		1 & 0 & 0
	\end{array} \right),
\end{eqnarray}
$M_{x,2} = (M_{x,1})^2$. Notice that sites have a vector space of dimension $d=2$ or $d=3$. In order to use a scale invariant MERA with $\chi > d$, we allow the disentanglers and isometries of the first few (typically one to five) layers to be different from $u$ and $w$. We iterate steps A1-A2 about 1000 times. With a cost per iteration that scales as $\chi^8$ and using a 3 GHz dual core desktop with 8 Gb of RAM, simulations for $\chi=4,8,16,22$ take of the order of minutes, hours, days and weeks respectively. The following results correspond to $\chi = 22$.

From Eq. \ref{eq:central} we obtain an estimate for the central charge, namely $c_{{\rm{Ising}}} = .5007$ and $c_{{\rm{Potts}}} = .806$, to be compared with the exact results $0.5$ and $0.8$. Fig. \ref{fig:CFTscaling} shows the smallest scaling dimensions $\Delta_{\alpha}$ of the scaling superoperator $\mathcal{S}$ \cite{ScDim}. We obtain remarkable agreement with those expected from CFT, as shown in %this table for various primary fields:
Table~\ref{tab:fdims}.
\begin{table}
%\begin{equation}
\begin{tabular}{|c|c|c|c|}
  \hline
  Ising & $\Delta^{\mbox{\tiny CFT}}$    &  $\Delta$ \scriptsize{(MERA $\chi=22$)}& rel. error \\ \hline
  $\sigma$ & 1/8 = 0.125 & 0.124997 & 0.002$\%$\\
  $\epsilon$ & 1 &  1.0001 & 0.01$\%$\\
  \hline
  Potts &  $\Delta^{\mbox{\tiny CFT}}$  & $\Delta$  \scriptsize{(MERA $\chi=22$)}& rel. error\\ \hline
  $\sigma_1$ & 2/15 = 0.1$\hat{3}$ & 0.1339 & 0.4$\%$ \\
  $\sigma_2$ & 2/15 = 0.1$\hat{3}$ & 0.1339 & 0.4$\%$\\
  $\epsilon$ & 4/5 = 0.8 &  0.8204 & 2.5$\%$\\
  $Z_1$ & 4/3 = 1.$\hat{3}$ &  1.3346 & 0.1$\%$\\
  $Z_2$ & 4/3 = 1.$\hat{3}$ &  1.3351 & 0.1$\%$\\
  \hline
  \end{tabular} \nonumber
%\end{equation}
\caption{\label{tab:fdims}Comparison of scaling dimensions of primary fields of the Ising and Potts models calculated using MERA ({$\Delta$\scriptsize{(MERA $\chi=22$)}}) with exact results known from CFT ({$\Delta^{\mbox{\tiny CFT}}$}).} % for the Ising and 3-level Potts models.}
\end{table}
Recall that all the critical exponents of the model can be obtained from the scaling dimensions of primary fields. For instance, for the Ising model the exponents $\nu$ and $\eta$ are $\nu = 2\Delta_{\sigma}$ and $\eta = \frac{1}{2-\Delta_{\epsilon}}$, whereas the \emph{scaling laws} express the critical exponents $\alpha,\beta,\gamma, \delta$ in terms of $\nu$ and $\eta$ \cite{Francesco}.
Further, the OPE coefficients for primary fields of, say, the critical Ising model are computed as follows. The matrix $C_{\alpha\beta}$ in Eq. \ref{eq:C2} is diagonal for the scaling operators corresponding to $\mathbb{I}$, $\sigma$ and $\epsilon$, which we normalize so that $C_{\alpha\beta} = \delta_{\alpha\beta}$. With this normalization, we then compute the coefficients $C_{\alpha\beta\gamma}$ using Eq. \ref{eq:C3}. We reproduce all the values of Eq. \ref{eq:OPE_Ising} with errors bounded by $3\times 10^{-4}$.

%\textbf{Discussion.---} In this paper we have explained how to compute the ground state of a critical Hamiltonian using the scale invariant MERA and how to extract from it the properties that characterize the system at the quantum critical point. Our results, which build upon those of Ref. \cite{ER,Free,MERA,Topological,Transfer,MERAalgorithmNEW}, also unveil a concise connection between the scale invariant MERA and CFT that adds significantly to the conceptual foundations of entanglement renormalization. Accordingly, the scale invariant MERA can be regarded as approximately realizing an infinite dimensional representation of the Virasoro algebra \cite{Cardy,Francesco}. The finite value of $\chi$ effectively implies that only a finite number of the quasi-primary fields of the theory can be included in the description. Fields with small scaling dimension, such as primary fields, are retained foremost. As a result, given a critical Hamiltonian on an infinite lattice, we can numerically estimate the scaling dimension and OPE of the primary fields of the CFT that describes the continuum limit of the model. 

\textbf{Discussion.---} In this paper we have explained how to compute the ground state of a critical Hamiltonian using the scale invariant MERA and how to extract from it the properties that characterize the system at a quantum critical point. Our results, which build upon those of Ref. \cite{ER,Free,MERA,Topological,Transfer,MERAalgorithmNEW}, also unveil a concise connection between the scale invariant MERA and CFT. This correspondence adds significantly to the conceptual foundations of entanglement renormalization. The scale invariant MERA can be regarded as approximately realizing an infinite dimensional representation of the Virasoro algebra \cite{Cardy,Francesco}. The finite value of $\chi$ effectively implies that only a finite number of the quasi-primary fields of the theory can be included in the description. Fields with small scaling dimension, such as primary fields, are retained foremost. 
%As a result, given a critical Hamiltonian on an infinite lattice, we can numerically estimate the scaling dimension and OPE of the primary fields of the CFT that describes the continuum limit of the model. 
As a result, given a Hamiltonian on an infinite lattice, we can numerically evaluate the scaling dimensions and OPE of the primary fields of the CFT that describes the continuum limit of the model. This approach differs in a fundamental way from, and offer an alternative to, the long-established techniques of Refs. \cite{CardyAll}, based instead on finite size scaling.
 
We conclude by noting that most of our considerations rely on scale invariance alone and can be applied to study also critical ground states in 2D systems \cite{2DMERA}.

We thank M. Aguado, L. Tagliacozzo, and W.-L. Yang for useful discussions. Support from the Australian Research Council (APA, FF0668731, DP0878830) is acknowledged.


\begin{thebibliography}{99}
\bibitem{clarify} In Ref. \cite{Transfer} the ascending/descending/scaling superoperators were called quantum MERA channel/MERA transfer matrix. We emphasize that these superoperators, an essential part of the MERA formalism, were introduced in Ref. \cite{MERA} (see e.g. Fig. 5 of Ref. \cite{MERA} for an explicit construction of the descending superoperator, where it is also evident that it is a quantum channel).
\bibitem{ScDim} Our numerics show that the lowest $n_{\Delta}$ scaling dimensions fulfill $\Delta^{(1)}_{\alpha}\approx \Delta_{\alpha} \approx \Delta_{\alpha}^{\mbox{\tiny CFT}}$, where $n_{\Delta}$ grows with $\chi$.
\bibitem{truncation} In practice we only compute the first $k$ terms ($k \approx 2,3$) of the expansion $\bar{h} = h_0 + h_1/3 + h_2/9 + \cdots$. This average is only needed when $H$ contains operators that are irrelevant in the RG sense.
\end{thebibliography}
\end{document}